\documentclass[preprint,showpacs,preprintnumbers,amsmath,amssymb]{revtex4}
\usepackage{graphicx}
\usepackage{dcolumn}
\usepackage{bm}

\newcommand{\be}{\begin{equation}}
\newcommand{\en}{\end{equation}}
\newcommand{\bea}{\begin{eqnarray}}
\newcommand{\ena}{\end{eqnarray}}

\begin{document}


\title{ Warm-Logamediate  inflationary universe model  }

\author{Ram\'on Herrera}
\email{ramon.herrera@ucv.cl} \affiliation{ Instituto de
F\'{\i}sica, Pontificia Universidad Cat\'{o}lica de
Valpara\'{\i}so, Casilla 4059, Valpara\'{\i}so, Chile.}
\author{Marco Olivares}
\email{marco.olivares@ucv.cl} \affiliation{ Instituto de
F\'{\i}sica, Pontificia Universidad Cat\'{o}lica de
Valpara\'{\i}so, Casilla 4059, Valpara\'{\i}so, Chile.}

\date{\today}

\begin{abstract}
  Warm inflationary universe models in the context of  logamediate
 expansion  are studied.
 General  conditions required for these models
 to be realizable  and
 discussed. This study is done in the weak and strong dissipative
 regimes.  The parameters of our models are constrained from the
 observational   data.
\end{abstract}

\maketitle

\section{Introduction}

In modern cosmology our notions concerning the early universe have
introduced a new element, the inflationary scenario of the
universe\cite{R1}, which provides an attractive epoch for solving
some of the problems of the standard big bang model, like the
flatness,  horizon etc..It is well known  Inflation can provide an
elegant mechanism to explain the large-scale structure \cite{R2}
and  a causal interpretation of the origin of the observed
anisotropy of the cosmic microwave background (CMB)
radiation\cite{astro,astro2}.

 On the other hand, it is well  know that warm inflation, as
opposed to the conventional cool inflation,  has the attractive
feature that it avoids the reheating period at the end of the
accelerated expansion because of the decay of the inflaton field
into radiation and particles during the slow roll phase
\cite{warm}. During the warm scenario the dissipative effects are
important, so that radiation production occurs concurrently
together with the inflationary expansion. The dissipating effect
arises from a friction term which describes the processes of the
scalar field dissipating into a thermal bath.    This scenario has
additional advantages, namely: i) the thermal fluctuations during
inflation may play a dominant role in producing the initial
fluctuations necessary for Large-Scale Structure (LSS) formation.
These density fluctuations arise from thermal rather than quantum
fluctuations \cite{62526,1126}, ii) the slow-roll conditions  can
be satisfied for steeper scalar potentials, iii)  the universe
stops inflating and "smoothly" enters in a radiation dominated
Big-Bang phase\cite{warm,taylorberera}, iv) it may contribute a
very interesting mechanism for baryogenesis, where the spontaneous
baryo/leptogenesis can easily be realized in this scenario
\cite{Bra}.  Recently, a new mechanism with a baryon asymmetry
from dissipative effects during  warm baryogenesis was studied in
Ref.\cite{BasteroGil:2011cx}.

 In this paper, we will examine warm inflation and the effective potentials  that
arise to study logamediate inflationary universe model where the dissipative effect is present. The logamediate model
was originally examined by
Barrow in Ref.\cite{R11}. During  logamediate inflation,  the
scale factor $a(t)=\exp(A[\ln t]^\lambda)$, with $A>0$ and
$\lambda>1$ constants. In particular, for $\lambda=1$, this scale
factor reduces to power law inflation i.e. $a\sim t^A$ \cite{atp}.
This accelerate expansion of the universe has the interesting
property that the ratio of tensor to scalar perturbations is small
and the power spectrum can be either red or blue tilted, according
to the values of the parameters appearing in the model \cite{R12}.

On the other hand, the condition for inflation to occur is that
the inflaton field slow-roll near the top of the potential for
sufficiently long time, so that the vacuum energy drives the
inflationary expansion of the Universe. Many models of inflation
have been proposed , based on single field or multi-field
potentials see e.g. Refs.\cite{R9,R10}. We found that for the case
of the weak dissipative regime, the field potential is
$V(\phi)\propto\,\phi^{\alpha}\;\exp [-\beta\,\phi^{\gamma}]$ and
this kind of potential coincides with the effective potential
found by Barrow in Ref.\cite{R11} . In particular, when $\alpha=0$
this $V(\phi)$  includes exponential potential that appears in
Kaluza-Klein theories, supergravity, and in super-string models
(see Ref.\cite{P1}). When $\beta=0$ it potential includes
power-law potentials\cite{P2}. In Ref.\cite{A2} was considered the
behavior  of the parameters $\gamma$ and $\alpha$ in all regions
of the  parameter space. We also find this sort of potentials  in
higher-dimensional theories, scalar-tensor theories, and
supergravity corrections \cite{P3}.  Also, in the context of the
curvaton reheating for this potential was studied in
Ref.\cite{Ramon} and also the form of this potential was used in
dark energy models\cite{P6}. Also, we found that for the case of
the strong dissipative regime, the scalar field potential
$V(\phi)$ is a function of the inverse gamma function
$\gamma_{\lambda}^{-1}\left[\frac{1}{\alpha_1}\;\ln\phi\right]$.
This sort of potentials was analyzed  in Refs.\cite{new}, for two
models of nonlocal scalar fields on cosmological backgrounds (see
also, Ref.\cite{new2}).



Thus, our aim in this paper is to analyze  an evolving logamediate
scale factor in the warm inflationary universe scenario. We will
do this for two regimes; the weak and the strong dissipative
regimes.

 The outline of the paper goes as follow: The next section
 presents a short description of  model. In the
sections III and VI,   we discuss the weak and strong dissipative
regimes in the logamediate scenario.  In both sections, we give
explicit expressions for the dissipative coefficient, the scalar
potential, the scalar power spectrum and the tensor-scalar ratio
for these models. At the end, section V exhibits our conclusions.
We chose units so that $c=\hbar=1$.

\section{The  Warm-Intermediate Inflationary phase.\label{secti} }

We start by considering   the flat Friedmann equation, by using
the Friedmann-Robertson Walker (FRW) metric, where
\begin{equation}
H^2=\frac{\kappa}{3}\,\rho=\frac{\kappa}{3}\,[\rho_{\phi}+\rho_\gamma],
\label{HC}
\end{equation}
here, $\rho=\rho_\phi+\rho_\gamma$ represents the total energy
density and the universe is filled with a self-interacting scalar
field of energy density $\rho_\phi$ and a radiation field with
energy density $\rho_\gamma$. The parameter $H=\dot{a}/a$ denotes
the Hubble parameter,  $a$ is the scale factor and $\kappa=8\pi
G=8\pi/m_p^2$, where  $m_p$ represents the Planck mass. In the
following, also we will assume that the energy density associated
to the scalar field is $\rho_\phi=\dot{\phi}^2/2+V(\phi)$ and
$V(\phi)=V$ is the scalar potential.

The dynamics equations for $\rho_\phi$ and $\rho_\gamma$ in the
warm inflationary scenario is described by\cite{warm}
 \be\dot{\rho_\phi}+3\,H\,(\rho_\phi+P_\phi)=-\Gamma\;\;\dot{\phi}^2, \label{key_01}
 \en
and \be \dot{\rho}_\gamma+4H\rho_\gamma=\Gamma\dot{\phi}^2
.\label{3}\en Here $\Gamma>0$ is the dissipation coefficient and
it is responsible of the decay of the scalar field into radiation
during the inflationary era. This coefficient can be assumed to be
a constant or a function of the scalar field $\phi$, or the
temperature $T$, or both \cite{warm}.  Dots mean derivatives with
respect to time.

Following, Refs.\cite{warm,62526} in the inflationary scenario the
energy density associated to the scalar field $\rho_\phi$
dominates over the energy density associated to the radiation
field $\rho_\gamma$ and the Eq.(\ref{HC}) becomes
\begin{eqnarray}
H^2\approx\frac{\kappa}{3}\,\rho_\phi.\label{inf2}
\end{eqnarray}
We also consider that  during  this scenario the radiation
production is quasi-stable\cite{warm,62526}, i.e.
$\dot{\rho}_\gamma\ll 4 H\rho_\gamma$ and $
\dot{\rho}_\gamma\ll\Gamma\dot{\phi}^2$.  From Eq.(\ref{3}) we
obtained that the energy density of the radiation field is given
by
 \begin{equation}
\rho_\gamma=\frac{\Gamma\dot{\phi}^2}{4H}=-\frac{\Gamma\,\dot{H}}{2\,\kappa\,H\,(1+R)},\label{rh}
\end{equation}
where $R$ is the rate defined as
\begin{equation}
 R=\frac{\Gamma}{3H },\label{rG}
\end{equation}
for the case of the  weak (strong) dissipation  regime, we have
$R< 1$ ($R> 1$). Here, we noted that the energy density of the
radiation field $\rho_\gamma$,  could be written as $\rho_\gamma=
C_\gamma\, T^4$, where $C_\gamma=\pi^2\,g_*/30$,  $g_*$ is the
number of relativistic degrees of freedom and $T$ is the
temperature of the thermal bath.

Considering  Eqs. (\ref{key_01}) and (\ref{inf2}), we can write
\begin{equation}
 \dot{\phi}^2= -\frac{2\,\dot{H}}{\kappa\,(1+R)},\label{inf3}
\end{equation}
and from Eqs.(\ref{rh}) and (\ref{inf3}), the temperature of the
thermal bath can be rewritten  as
\begin{equation}
T= \left[-\frac{\Gamma\,\dot{H}}{2\,\kappa\,\,C_\gamma
H\,(1+R)}\right]^{1/4}.\label{rh-1}
\end{equation}

On the other hand, the dissipation coefficient  is computed for
models in cases of low-temperature regimes in Refs.\cite{26,28}.
This dissipation coefficient, was developed in supersymmetric
models which have an inflaton together with multiplets of heavy
and light fields. Here, three super-fields $\Phi$, $X$ and $Y$
with a super-potential, $W=g\Phi X^2+hXY^2$, where $\phi$, $\chi$
and $y$ refer to their bosonic component were study. The inflaton
field couples to heavy bosonic field $\chi$ and fermions
$\psi_\chi$, obtain their masses through couplings to $\phi$,
where $m_{\psi_\chi}=m_\chi=g\phi$.  Following Ref.\cite{26}, the
transfer of energy from the inflaton the mass $m_\phi$ to another
fields is dependent the coupling strength of the interaction, the
relative sizes of the mass $m_\phi$, the mass of decay products
and the temperature.  The scenario of concern occurs  when
 $m_\chi,m_{\psi_\chi}>T>H$ and $m_y,m_{\psi_y}<T$.  These
 conditions specifies to here as the low-temperature scenario \cite{26}.
Considering this scenario, the coefficient $\Gamma$, when $X$ and
$Y$ are singlets is given by
\begin{equation}
\Gamma\simeq0.64\,g^2\,h^4\left(\frac{g\,\phi}{m_\chi}\right)^4\,
\frac{T^3}{m_\chi^2}=C_\phi\,\frac{T^3}{\phi^2},\label{G}
\end{equation}
where $C_\phi=0.64\,h^4\,\cal{N}$ in which
${\cal{N}}={\cal{N}}_\chi {\cal{N}}_{decay}^2$. Here,
$\cal{N}_\chi$ is the multiplicity of the $X$ superfield and
${\cal{N}}_{decay}$ is the number of decay channels available in
$X$'s decay\cite{26,27}  (see also,
Refs.\cite{Berera:2008ar,BasteroGil:2010pb}).

From Eqs.(\ref{rh-1}) and  (\ref{G}) we get
\begin{equation}
\Gamma^{1/4}\,(1+R)^{3/4}\simeq \,\frac{\alpha_{0}}{\phi^2}
\left[-\frac{\dot{H}}{H}\right]^{3/4},\label{newG1}
\end{equation}
and so
\begin{equation}
\Gamma\,\simeq \,\frac{\alpha_{0}^4}{\phi^8}
\left[-\frac{\dot{H}}{H\,(1+R)}\right]^{3},\label{G1}
\end{equation}
where $ \alpha_{0}\equiv C_\phi\left[\frac{1}{2\kappa\,
C_\gamma}\right]^{3/4}$. The Eq.(\ref{G1}) determines the
dissipation coefficient in the weak (or strong) dissipative regime
in terms of scalar field  and the parameters of the model.

Considering  Eqs.(\ref{HC}) and (\ref{rh}) the scalar potential
becomes

\begin{equation}
V(\phi)=\frac{1}{\kappa}\left[3\,H^2+\frac{\dot{H}}{(1+R)}\,\left(1+\frac{3}{2}\,R\right)\right],\label{pot}
\end{equation}
which could be expressed explicitly in terms of the scalar field,
in the  weak (or strong) dissipative regime.

Following Ref.\cite{R12}, a generalized model of expansion of the
universe is logamediate inflation. Here,  the scale factor $a(t)$,
is given by
\begin{equation} a=\exp[\,A\,(\ln t)^{\lambda}],\label{at}
\end{equation}
where  $\lambda$ and $A$ are dimensionless constants parameters
such that $\lambda > 1$, and $A > 0$, see Ref.\cite{R12}.
Recalling, that for $\lambda=1$ and $A=p$, the logamediate
inflation model becomes power-law inflation, in which $a\sim t^p$.

 In the following, we study  models for
a variable dissipation coefficient $\Gamma$, and we will restrict
ourselves to the weak (or strong ) dissipation regime.

\section{ The  weak dissipative regime.\label{section2}}

Assuming that, once the system evolves according to the weak
dissipative regime, i.e. $\Gamma<3H$ and considering
Eqs.(\ref{inf3}) and (\ref{at}), we obtained a relation between
the scalar field and cosmological times given by
\begin{equation}
\phi(t)=\phi_0+\sqrt{\frac{2\,A\,\lambda}{\kappa}}\left[\frac{2}{1+\lambda}\right](\ln
t)^{\frac{1+\lambda}{2}},\label{wr1}
\end{equation}
where $\phi_0$ is  constant.  The Hubble parameter as a function
of the inflaton field,  becomes
\begin{equation}
H(\phi)=(A\,\lambda)B^{\lambda-1}\phi^{\gamma(\lambda-1)}\exp[-B\,\phi^{\gamma}],\label{HH}
\end{equation}
where
$$
\gamma\equiv\frac{2}{\lambda+1}\,\;\;B\equiv\left[\frac{1}{\gamma}
\sqrt{\frac{\kappa}{2A\lambda}}\right]^{\gamma}.
$$
Without loss of generality the constant $\phi_0=0$.

From Eq.(\ref{pot}) the scalar potential in this regime is
$V=(3H^2+\dot{H})/\kappa$.  At late times and following
Ref.\cite{R12}, the scalar potential  becomes
\begin{equation}
V(\phi)=V_{0}\phi^{\alpha}\exp[-\beta\,\phi^{\gamma}],\label{pot11}
\end{equation}
where $ V_{0}=\frac{3}{\kappa}(A\lambda)^{2}B^{2(\lambda-1)},
\,\;\;\alpha=2\gamma(\lambda-1)$ and $\beta=2B$.

Note that this kind of potential coincides with the scalar
potential found in Ref.\cite{R12}.  Note also that the scalar
field $\phi$, the Hubble parameter $H$, and the potential
$V(\phi)$ become independent of the parameters $C_\phi$ and
$C_\gamma$ in this regime.

From Eq.(\ref{G1})  the dissipation coefficient $\Gamma$ as
function of scalar field $\phi$, results
\begin{equation}
\Gamma(\phi)= \alpha_{0}^{4}\,\phi^{-8}\exp[-3B\,\phi^{\gamma}].
\end{equation}

Considering, the dimensionless slow-roll parameter $\varepsilon$,
we get
\begin{equation}
\varepsilon=-\frac{\dot{H}}{H^2}=(A\lambda)^{-1}B^{-(\lambda-1)}\phi^{-\gamma(\lambda-1)},\label{ep}
\end{equation}
and the other slow-roll parameter  $\eta$, becomes
\begin{equation}
\eta=-\frac{\ddot{H}}{H \dot{H}}=(A\lambda
\,B^{\lambda})^{-1}\phi^{-\gamma\lambda}\left[2B\phi^{\lambda}-(\lambda-1)\right]\,.\label{eta}
\end{equation}

So, the condition for inflation to occur  $\ddot{a}>0$ (or
equivalently $\varepsilon<$1)   is only satisfied when
$\phi>[A\lambda B^{(\lambda-1)}]^{\frac{-1}{\gamma(\lambda-1)}}$.

The number of e-folds $N$ between two different values of
cosmological times $t_1$ and $t_2$ or equivalently between two
values of the scalar field $\phi_1$ and $\phi_2$, from
Eq.(\ref{wr1})   results
\begin{equation}
N=\int_{t_1}^{t_{2}}\,H\,dt=A\,\left[(\ln t_{2})^{\lambda}-(\ln
t_{1})^{\lambda}\right]=A
B^{\lambda}\,\left(\phi^{\gamma\lambda}_{2}-\phi^{\gamma\lambda}_{1}\right).\label{N1}
\end{equation}

Following Ref.\cite{R12} and  considering that inflation begins at
the earliest possible stage, where $\varepsilon=1$ or equivalently
$\ddot{a}=0$, the scalar field $\phi_1$, is given by
\begin{equation}
\phi_{1}=[A\lambda
B^{(\lambda-1)}]^{\frac{-1}{\gamma(\lambda-1)}}\;. \label{al}
\end{equation}

On the other hand, the density perturbation could be written as
${\cal{P}_{\cal{R}}}^{1/2}=\frac{H}{\dot{\phi}}\,\delta\phi$\cite{warm}.
In particular in the warm inflation regime, a thermalized
radiation component is present, therefore, inflation fluctuations
are dominantly thermal rather than quantum\cite{warm,62526}.  In
the weak dissipation limit, we have $\delta\phi^2\simeq H\,T$
\cite{62526,B1}. From Eqs.(\ref{inf3}) and (\ref{rh-1}), the
density perturbation is given by
\begin{equation}
{\cal{P}_{\cal{R}}}=
\beta_4\;\phi^{\alpha-2}\exp[-\beta\,\phi^{\gamma}], \label{pd}
\end{equation}
where
$$
\beta_4=\frac{1}{4}\left(\frac{C_{\phi}}{C_{\gamma}}\right)(A\lambda)^{2}B^{2(\lambda-1)}.
$$
The scalar spectral index $n_s$ is given by $ n_s -1 =\frac{d
\ln\,{\cal{P}_R}}{d \ln k}$. Here,  the interval in wave number
$k$ is related to the number of e-folds $N$, through $d \ln
k(\phi)=d N(\phi)=(H/\dot{\phi})\,d\phi$. From Eqs. (\ref{wr1})
and (\ref{pd}), we get
\begin{equation}
n_s=1-\frac{2B^{-(\lambda-1)}}{A\,\lambda}\phi^{-\gamma(\lambda-1)}.\label{nss1}
\end{equation}
We noted that the  index $n_s$ can be re-expressed in terms of the
number  $N$. Combining  Eqs.(\ref{N1}) and (\ref{al}) we obtain
\begin{equation}
n_s=1-\frac{2}{A\,\lambda}\left[\frac{N}{A}+(A\,\lambda)^{\frac{-\lambda}{\lambda-1}}\right]^{-\frac{\lambda-1}{\lambda}}.
\end{equation}

\begin{figure}[th]
\includegraphics[width=3.5in,angle=0,clip=true]{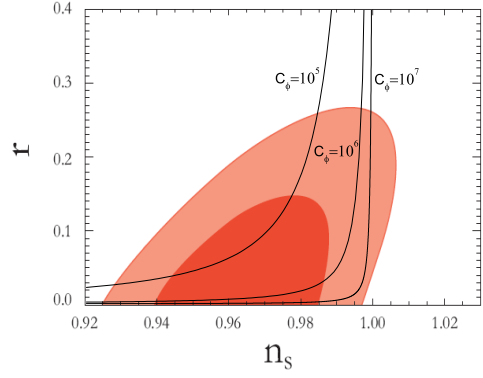}
\includegraphics[width=3.5in,angle=0,clip=true]{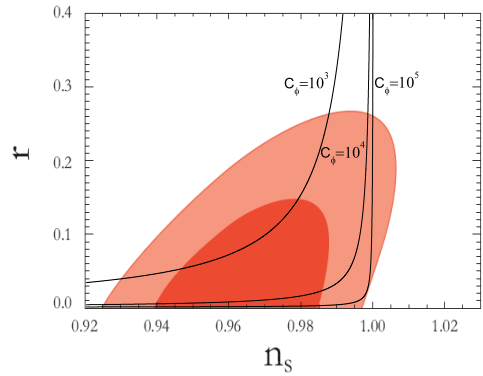}
\caption{Evolution of the tensor-scalar ratio $r$ versus the
scalar spectrum index $n_s$ in the weak dissipative regime, for
three different values of the parameter $C_\phi$. In the upper
panel, we have take $\lambda=5$ and  the lower panel,
$\lambda=20$. In both panels  we used $A=10^{-1}$, $\kappa=1$ and
$C_\gamma=70$. \label{rons}}
\end{figure}

As it was remarked in Ref.\cite{Bha} the generation of tensor
perturbations during inflation would  produce  gravitational wave.
The corresponding spectrum  is given by
${\cal{P}}_g=8\kappa(H/2\pi)^2$. For this regime and from
Eq.(\ref{pd}) we may write the tensor-scalar ratio $r$ as
\begin{equation}
r(k)=\left(\frac{{\cal{P}}_g}{P_{\cal R}}\right) \simeq
\beta_5\;\phi^2, \label{Rk}\end{equation} where $
\beta_5=\frac{8\kappa \,C_{\gamma}}{\pi^{2}\,C_{\phi}} $.

The tensor to scalar ratio can be written in terms of the scalar
spectral index $n_s$ as
\begin{equation}
r\simeq \beta_5
\left[\frac{2B^{-(\lambda-1)}}{A\,\lambda(1-n_s)}\right]^{\frac{2}{\gamma(\lambda-1)}}.
\label{Rk1}
\end{equation}

Analogously, the tensor to scalar ratio as function of the number
of e-folding $N$ results as

\begin{equation}
r
\simeq\beta_5\,B^{-2/\gamma}\left[\frac{N}{A}+(A\,\lambda)^{\frac{-\lambda}{\lambda-1}}\right]^{\frac{2}{\lambda\,\gamma}}.
\label{Rk11}\end{equation}

In Fig.(\ref{rons}) we shows the dependence of the tensor-scalar
ratio on the spectral index (for the weak regime) for the special
case in which we fixe $A=10^{-1}$, and we have used three
different values of the parameter $C_\phi$. In the upper panel we
used $\lambda=5$ and in the lower panel $\lambda=20$.

The seven-year WMAP data places stronger limits on the tensor to
scalar ratio $r$. In order to write down values that relate $n_s$
and $r$, we used Eq.(\ref{Rk1}). Also in both panels we have used
the values $C_\gamma=70$ and $\kappa=1$. From the upper panel in
which $\lambda=5$, we noted that for the value of the parameter
$C_\phi>10^5$, the model is well supported by the data in the weak
regime. Analogously, from de lower panel, we noted that for
$C_\phi>10^3$, the model is well supported by the data. In
general, we noted that when we increases the values of  $\lambda$,
the values of the parameters $C_\phi$ decreases (see
Fig.(\ref{rons})).  If we compared with respect to monomial
potentials, the value of the  parameter $C_\phi$, increase by two
orders of magnitude and for the Hybrid models, $C_\phi$ increase
by one order of magnitude,  when $\lambda=5$
\cite{BasteroGil:2009ec}.


\section{ The  strong dissipative regime.\label{section3}}

We consider now the case in which  $R=\Gamma/3H>1$. From
Eqs.(\ref{inf3}) and (\ref{at}), we can obtained a relation
between the scalar field and cosmological times given by
\begin{equation}
\ln\left[\phi(t)/\phi_0\right]=\,\alpha_1\,
\gamma_{\lambda}(t),\label{wr12}
\end{equation}

where $\phi_0$ is a constant (without loss of generality we can
taken $\phi_0=1$), $\alpha_1$ is defined by
$$
\alpha_1=\sqrt{\frac{2\cdot3^{1/4}}{\kappa\alpha_{0}}}(A\,\lambda)^{5/8}|(-4)^{1+\frac{5}{8}(\lambda-1)}|,\;\;\;\mbox{and}
\,\,\;\gamma_{\lambda}(t)\equiv|\gamma[1+\frac{5}{8}(\lambda-1),-\frac{1}{4}\ln
t]|.
$$
Here, $\gamma_{\lambda}(t)$ is the incomplete gamma function, see
e.g. Ref.\cite{Libro}.


The Hubble parameter as a function of the inflaton field, $\phi$,
is given by
\begin{equation}
H(\phi)=A\,\lambda\,\left(\gamma_{\lambda}^{-1}\left[\frac{1}{\alpha_1}\;\ln\phi\right]\right)^{-1}
\left(\ln\gamma_{\lambda}^{-1}\left[\frac{1}{\alpha_1}\;\ln\phi\right]\right)^{\lambda-1},\label{HH2}
\end{equation}
where
$\gamma_{\lambda}^{-1}\left[\frac{1}{\alpha_1}\;\ln\phi\right]$ is
the inverse gamma function of $\gamma_{\lambda}(t)$.

At the last times, analogously to the case of the weak dissipative
regime,   the scalar potential from Eq.(\ref{pot}), becomes

\begin{equation}
V(\phi)=
\frac{K}{R^2\,\phi^4}\left(\gamma_{\lambda}^{-1}\left[\frac{1}{\alpha_1}
\;\ln\phi\right]\right)^{-3}
\left(\ln\gamma_{\lambda}^{-1}\left[\frac{1}{\alpha_1}\;\ln\phi\right]\right)^{3(\lambda-1)/2}
,\label{pot11}
\end{equation}
where $K=3^{1/2}\,\alpha_0^2\,(A\,\lambda)^{3/2}/\kappa$.


Considering Eq.(\ref{G1}) the dissipation coefficient,  can be
expressed as a function of the scalar field  as
\begin{equation}
\Gamma(\phi)=(3A\lambda)^{3/4}\alpha_{0}\phi^{-2}\left(\gamma_{\lambda}^{-1}\left[\frac{1}{\alpha_1}
\;\ln\phi\right]\right)^{-3/2}
\left(\ln\gamma_{\lambda}^{-1}\left[\frac{1}{\alpha_1}\;\ln\phi\right]\right)^{\frac{3}{4}(\lambda-1)}.\label{gg2}
\end{equation}

The dimensionless slow-roll parameter $\varepsilon$ for this
regime, becomes
\begin{equation}
\varepsilon=-\frac{\dot{H}}{H^2}=(A\lambda)^{-1}\left(\ln\gamma_{\lambda}^{-1}\left[\frac{1
}{\alpha_{1}}\ln \phi\right]\right)^{-(\lambda-1)},\label{ep1}
\end{equation}
and the slow-roll parameter  $\eta$, is given by
\begin{equation}
\eta=-\frac{\ddot{H}}{H
\dot{H}}=(A\lambda)^{-1}\left(\ln\gamma_{\lambda}^{-1}\left[\frac{1
}{\alpha_{1}}\ln
\phi\right]\right)^{-\lambda}\left\{2\left(\ln\gamma_{\lambda}^{-1}\left[\frac{1
}{\alpha_{1}}\ln
\phi\right]\right)-(\lambda-1)\right\}.\label{eta2}
\end{equation}

Again, following Ref.\cite{R12}  the condition $\varepsilon=1$ at
the beginning of inflation the scalar field, results
\begin{equation}
\phi_{1}=\exp\left[\alpha_1\;\gamma_{\lambda}\left(e^{\mu}\right)\right],
\label{al22}
\end{equation}
where $ \mu\equiv(A\lambda)^{\frac{-1}{\lambda-1}}. $

The number of e-folds $N$ in this regime, from Eq.(\ref{wr12}) is
given by
\begin{equation}
N=\int_{t_1}^{t_{2}}\,H\,dt=A\,
\left\{\left(\ln\gamma_{\lambda}^{-1}\left[\frac{1}{\alpha_1}\;\ln\phi_2\right]\right)^{\lambda}
-\left(\ln\gamma_{\lambda}^{-1}\left[\frac{1}{\alpha_1}\;\ln\phi_1\right]\right)^{\lambda}\right\}.
\label{N22}
\end{equation}

In the case of high dissipation regime and following
Ref.\cite{Bere2}, we can write
$\delta\phi^2\simeq\,\frac{k_F\,T\,}{2\,\pi^2}$, where  the
wave-number  $k_F=\sqrt{\Gamma H}=H\,\sqrt{3 R}> H$, and
corresponds to the freeze-out scale at which dissipation damps out
to the thermally excited fluctuations.
From Eqs.(\ref{wr12}) and (\ref{gg2}) we obtained that
\begin{equation}
{\cal{P}_{\cal{R}}}=
\alpha_2\,\phi^{-3}\,\left(\gamma_{\lambda}^{-1}\left[\frac{1}{\alpha_{1}}\ln
\phi\right]\right)^{-9/4}
\left(\ln\gamma_{\lambda}^{-1}\left[\frac{1}{\alpha_{1}}\ln
\phi\right]\right)^{\frac{15}{8}(\lambda-1)}, \label{pd21}
\end{equation}
where
$$
\alpha_2=3^{3/8}\alpha_{0}^{3/2}(4\pi^{2})^{-1}\left(\frac{\kappa^{3}}{2C_{\gamma}}\right)^{1/4}(A\lambda)^{15/8}.
$$
Considering  Eqs.(\ref{wr12}) and (\ref{gg2}) the scalar spectral
index $n_s=d\,{\cal{P}_{\cal{R}}}/d\ln k$, is given by
\begin{equation}
n_s=
1-\frac{9}{4A\lambda}\left(\ln\gamma_{\lambda}^{-1}\left[\frac{1}{\alpha_1}\;\ln\phi\right]\right)^{-(\lambda-1)}.\label{nss}
\end{equation}
Analogously as the weak regime, the scalar spectra index,  can be
write in terms of the number of e-folds. Considering Eqs.
(\ref{al22}) and (\ref{N22}) results
\begin{equation}
n_s=
1-\frac{9}{4A\,\lambda}\left[\frac{N}{A}+(A\,\lambda)^{\frac{-\lambda}{\lambda-1}}\right]^{-\frac{\lambda-1}{\lambda}}.
\label{ns2}
\end{equation}

For this regime  we may write the tensor-scalar ratio, using Eqs.
(\ref{HH2}) and (\ref{pd21}), we get
\begin{equation}
r(k)=\left(\frac{{\cal{P}}_g}{P_{\cal R}}\right)
=\,\alpha_{3}\phi^{3}\left(\gamma_{\lambda}^{-1}\left[\frac{1}{\alpha_1}\;\ln\phi\right]\right)
^{1/4}\left(\ln\gamma_{\lambda}^{-1}\left[\frac{1}{\alpha_1}\;\ln\phi\right]\right)^{\frac{\lambda-1}{8}},
\label{Rk2}
\end{equation}
where $ \alpha_3=\frac{2\kappa(A\lambda)^{2}}{\pi^{2}\alpha_{2}}$.

 Analogously as the weak regimen, the tensor to scalar ratio
$r$, in terms of the scalar spectral index, becomes
\begin{equation}
r=
\alpha_3(F[n_{s}])^{\frac{\lambda-1}{8}}\exp[\frac{F[n_{s}]}{4}+3\alpha_1
\gamma_{\lambda}(e^{F[n_{s}]})], \label{Rk21}
\end{equation}
where
$$
F[n_{s}]\equiv\left[\frac{9}{4A\lambda(1-n_s)}\right]^{\frac{1}{\lambda-1}}.
$$

Also, we can write the tensor-scalar ratio as function of the
number of e-folding

\begin{equation}
r
=\alpha_3(G[N])^{\frac{\lambda-1}{8}}\exp[\frac{G[N]}{4}+3\alpha_1
\gamma_{\lambda}(e^{G[N]})], \label{Rk22}
\end{equation}

where
$
G[N]\equiv\left[\frac{N}{A}+(A\,\lambda)^{\frac{-\lambda}{\lambda-1}}\right]^{1/\lambda}.
$

\begin{figure}[th]
\includegraphics[width=3.5in,angle=0,clip=true]{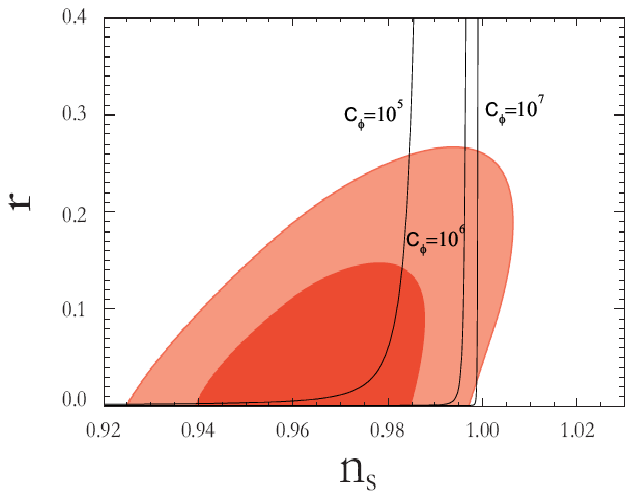}
\includegraphics[width=3.5in,angle=0,clip=true]{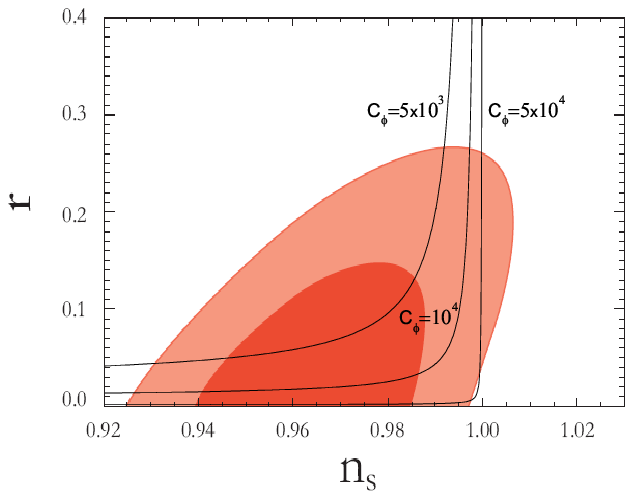}
\caption{Evolution of the tensor-scalar ratio $r$ versus the
scalar spectrum index $n_s$ in the strong dissipative regime, for
three different values of the parameter $C_\phi$. In the upper
panel we have used $\lambda=5$ and the lower panel $\lambda=20$.
In both panels  we used $A=10^{-1}$, $\kappa=1$ and $C_\gamma=70$.
\label{fig2}}
\end{figure}


For the strong dissipative regime i.e., $R=\Gamma/3H>1$, the
Fig.(\ref{fig2})  shows  the dependence of the tensor-scalar ratio
on the spectral index for the special case in which we fixe
$A=10^{-1}$, and we have used three different values of the
parameter $C_\phi$. In the upper panel we used $\lambda=5$ and in
the lower panel $\lambda=20$.

In order to write down values that relate $n_s$ and $r$, we used
Eq.(\ref{Rk21}). Also in both panels we have used the values
$C_\gamma=70$ and $\kappa=1$. From the upper panel in which
$\lambda=5$, we noted that for the value of the parameter
$C_\phi>10^4$, the model is well supported by the data in the weak
regime. Analogously, from de lower panel in which $\lambda=20$, we
noted that for $C_\phi>10^2$, the model is well supported by the
data. Analogous as the weak regime,  we noted that when we
increases the values of $\lambda$, the values of the parameter
$C_\phi$ decreases. Analogous as the weak scenario, if we compared
with respect to monomial potentials, the value of the parameter
$C_\phi$ increase by three orders of magnitude and  for the Hybrid
models, $C_\phi$ increase by two orders of magnitude
\cite{BasteroGil:2009ec}, when $\lambda=5$.

\section{Conclusions \label{conclu}}

In this paper we have studied the warm-logamediate inflationary
model in the weak and  strong dissipative regimes.  In the
slow-roll approximation we have found explicit expressions for the
corresponding dissipation parameter $\Gamma$, scalar potential
$V$, the number of e-folds $N$, power spectrum of the curvature
perturbations $P_{\cal R}$, tensor-scalar ratio $r$ and scalar
spectrum index $n_s$.

 When
$\Gamma<3H$ warm inflation occurs in the so-called weak
dissipative regime. In this case, the dissipation coefficient
$\Gamma\propto\phi^{-8}\exp[-3B\,\phi^{\gamma}]$ and the scalar
potentia $V(\phi)\propto\phi^{\alpha}\exp[-\beta\,\phi^{\gamma}]$
that coincides with Ref.\cite{R12}. In order to bring some
explicit results we have taken trajectories for different
combinations of the parameters in the   $r-n_s$ plane to
first-order in the slow roll approximation. We  noted that the
parameter $C_\phi$, which is bounded from bellow, $C_\phi>10^5$,
the model is well supported by the data as could be seen from
upper panel of Fig.(\ref{rons}) in which $\lambda=5$.We also noted
that for the lower panel where $\lambda=20$, the values of  the
parameter $C_\phi>10^{3}$ the model is well supported by the data.
In both panel, we have used the WMAP seven year data and also we
have taken the value $A=10^{-1}$, $\kappa=1$ and $C_\gamma=70$. On
the other hand, when $\Gamma>3H$ warm inflation occurs in the
so-called strong dissipative regime. In this regime, the effective
potential is given by Eq.(\ref{pot11}) and the dissipation
coefficient $\Gamma$ by Eq.(\ref{gg2}). Analogously as the weak
regime the Fig.(\ref{fig2}) shows  trajectories for different
combinations of the parameters $C_\phi$ in the   $r-n_s$ plane,
for $A=10^{-1}$ and $C_\gamma=70$. Curiously, in both regimes we
noted that when we increases the values of  $\lambda$, the values
of the parameter $C_\phi$ decreases.

In this paper, we have not addressed the non-Gaussian effects from
the non-linearity parameter $f_{NL}$, during warm-logamediate
inflation (see e.g., Refs.\cite{27,fNL}).  We hope to return to
this point in the near future.

\begin{acknowledgments}
R.H. was supported by COMISION NACIONAL DE CIENCIAS Y TECNOLOGIA
through FONDECYT grants  N$^0$ 1090613 and N$^0$ 1110230. M.O. was
supported by Proyecto D.I. PostDoctorado 2012 PUCV.
\end{acknowledgments}


\end{document}